# The Impact of Perceptions of Social Media Advertisements on Advertising Value, Brand Awareness and Brand Associations: Research on Generation Y Instagram Users

İbrahim Halil Efendioğlu[1]  and Yakup Durmaz[2]

### Abstract

*The aim of this paper is to examine how consumer perceptions of social media advertisements affect advertising value and brand awareness. With a rapid increase in the number of social media users over the last ten years, a new advertising domain has become available for companies. Brands that manage social media well in their advertising strategies can quickly influence consumer decision-making and create awareness. However, in social media advertising, which is different from traditional advertising, creating content should be produced and this content should be perceived in a short time by consumers. To achieve this, it is necessary to build rapport with consumers and to present correctly what they wish to see in advertisements by creating awareness. In view of the increasing importance of social media advertising, the study examines how consumer perceptions of Instagram advertisements affect advertising value and brand awareness. This study was conducted with Generation Y consumers on the basis of their Instagram habits, a popular social media app. For this purpose, surveys were held with 665 participants who use Instagram. The collected data were analyzed using structural equation modeling. According to the analysis results, Y-generation's perceptions of Instagram advertisements have both a positive and negative impact on advertising value and brand awareness and brand associations.*

**Keywords:** *Social Media Advertisements; Instagram Advertising; Consumer Perceptions; Advertising Value; Brand Awareness; Brand Associations*

## Introduction

Today 4.2 billion out of the 7.8 billion people worldwide actively use social media (We Are Social, 2022). Companies are investing more in social media marketing to promote their services or products (Baum et al., 2019). Companies take social media into account, in addition to traditional advertising domains when planning their advertising budgets. Advertising expenditures for social media are estimated to reach $173 billion in 2022 (Hootsuite, 2022). Although social media has become an integral part of everyday life, the results of social media marketing campaigns are sometimes unsatisfactory. This shows the reason why social media advertisements need to be properly designed with the creating content to appeal to consumers. However, it is necessary to measure how social media advertisements are perceived by the consumer and whether they have the desired effect on the consumer. Occasionally, some social media advertisements simply fail, as social media advertising is

---

[1] Dr. Ibrahim Halil Efendioglu, Department of Informatics, Gaziantep University, Gaziantep, Turkey.
E-mail: efendioglu@gantep.edu.tr
[2] Assoc. Prof. Dr. Yakup Durmaz, Department of Business Administration, Hasan Kalyoncu University, Gaziantep, Turkey.
E-mail: yakup.durmaz@hku.edu.tr
**Acknowledgement**: This article was produced from İbrahim Halil Efendioğlu's doctoral thesis.

inherently slightly different from traditional advertising channels (Dehghani et al., 2016; Duffett, 2015; Murillo et al., 2016; Shareef et al., 2019).

Advertising value is the subjective opinion of consumers about how useful and successful an advert is. When it is possible to measure how advertisements are perceived by consumers, insights can be gained into the effectiveness and overall success of an advertisement (Ducoffe, 1995). Ducoffe's advertising value model is often used in literature to solve this measurement issue.

Instagram has rapidly gained popularity in recent years. Instagram is an important social media platform that was launched on October 6, 2010. With the Instagram application, users take photos or videos of special moments in their lives and share them with their followers (Hu et al., 2014). About 100 million photos and videos are shared every day on this platform which attracts more and more users. After Facebook bought Instagram for $1 billion in 2012, Instagram has grown and changed significantly. The number of users rose from 80 million in 2012 to 2 billion in 2022 (Statista, 2022). Brands are expected to spend $10 billion on Instagram Advertising in 2022 (Kusumasondjaja and Tjiptono, 2019).

Consumers are often exposed to new services, products and new brands (Sirkeci and Zeren, 2018). With the increase in digital marketing, consumer perceptions have also changed (Bhagat and Rajan, 2021). After all, advertisements are about creating perceptions. Through such efforts, companies want consumers to perceive their brands differently than other brands. This is possible through advertising, especially in marketing communication (Erciş and Çat, 2016). Instagram has become, among Generation Y users, a popular platform whose popularity continues to rise (We Are Social, 2022). Millennials, born between 1981 and 2000, also have the largest share of the global online market. Advertisers and companies are increasingly focusing on millennials for Instagram (Jamie et al., 2021). Generation Y consumers with high purchasing power use Instagram to search for information before purchasing a product. This is why businesses are getting more exposure to Instagram ads (Aprilia and Setiadi, 2017). Therefore, the research was carried out among the Y generation.

On that basis, the main problem posited by the research is: "Does the perception of Instagram advertisements influence the advertising value and brand awareness from the point of view of Generation Y?" When all these are combined, social media advertisements that are quickly perceived and create awareness will be successful and effective. In this context, the departure point of this study is the effect of the perception of social media advertisements on advertising value and brand awareness. The objective of the study is therefore to determine the influence of perceptions of Instagram advertisements on advertising value and brand awareness amongst Generation Y. To do that, the study proposes a new conceptual model that combines the brand awareness and the advertising value model.

The research has been organized as follows. In the first part of the study, previous studies and related concepts are explained and discussed. This is followed by hypotheses that influence the research model, advertising value and brand awareness and the arguments supporting them. The conceptual framework of the research will then be defined. In the Methodology section, information will be provided about the data collection and analysis period. And in the final section, the theoretical and practical contributions of the study will be discussed.

## Literature review

Consumer perceptions and attitudes towards social media advertising were analyzed in different social media environments in literature. Such studies utilized different social media environments like Facebook (Erkan et al., 2019; Ertugan, 2017; Logan et al., 2012; Shareef et al., 2019; Turgut et al., 2016), YouTube (Dehghani et al., 2016; Şahin, 2018), Twitter (Murillo et al., 2016), and Instagram (Chen, 2018; Gaber et al., 2019; Sabuncuoğlu and Gülay, 2016; Zengin and Zengin, 2017). In addition, a number of studies have been conducted on the perception of adverts by different age groups and different generations (Cop and Baş, 2010; Elden et al., 2014; Sabuncuoğlu and Gülay, 2016; Ünal et al., 2011). There are also studies analyzing advertisements' impact on purchasing intent and behavior (Aksoy and Gür, 2018; Dehghani and Tumer, 2015; Dehghani et al., 2016; Erciş and Çat, 2016; Hutter et al., 2013; Martins et al., 2018; Turgut et al., 2016; Ünal et al., 2011; Wang et al., 2009) and on word-of-the-mouth advertising (Hutter et al., 2013; Kim and Ko, 2012; Turgut et al., 2016; Yazgan et al., 2014; Yıldız, 2015) . In studies examining consumers' attitudes towards advertisements in different media environments, for instance TV (Brackett and Carr, 2001; Ducoffe, 1995; Elden et al., 2014; Erciş and Çat, 2016; Khan et al., 2016; Logan et al., 2012; Özbucak Albar and Öksüz, 2013), or mobile environments (Bayrak Meydanoğlu, 2016; Liu et al., 2012; Martins et al., 2018; Ünal et al., 2011) consumer perceptions of advertisements were studied. In addition, various studies were conducted on brand awareness (Alemdar and Dirik, 2016; Bilgin, 2018; Hoyer and Brown, 1990; Khan et al., 2016; MacDonald and Sharp, 2000; Özbucak Albar and Öksüz, 2013; Romaniuk et al., 2004) to analyze the relationship between consumer preferences and the brand.

### *Advertising value*

Advertising value reflects consumer satisfaction and experience. Thus, it is the value of the advertising in the eyes of the consumer and tells whether the product meets customer expectations (Zeithaml, 1988). On social media, customers create content, share information and ideas, and thus interact with others. In this way, more active users can create new advertising values for businesses and consumers (Akyüz, 2013). However, advertisements with inappropriate content and unethical components cause a negative perception of value. So, situations may arise that negatively influence the attitudes of individuals (Balakrishnan et al., 2013). Therefore, companies should accurately identify the desired target audience and attach greater importance to advertising value to ensure that the messages sent through advertisements are well perceived (Akkaya et al., 2017). Advertising value is explained by four main components. These are informativeness, deceptiveness, entertainment, and irritation (Ducoffe, 1995).

### *Informativeness*

Advertising plays an important role in the transmission of information, as consumers want to quickly access the latest information. Advertisements are ideal for giving people information (Ling et al., 2010). Making informative advertisements with an up-to-date message highly benefits the consumer (Aydın, 2017). In addition, advertisements should be informative so that users can make informed decisions. In the end, such advertisements provide information about alternative options (Lee and Hong, 2016). Studies on the consumer perception of advertisements have shown that informative advertising has a positive influence on advertising value (Bayrak Meydanoğlu, 2016; Brackett and Carr, 2001; Dehghani et al., 2016;

Elden et al., 2014; Erkan et al., 2019; Kayapınar et al., 2017; Ling et al., 2010; Logan et al., 2012; Liu et al., 2012; Martins et al., 2018; Murillo et al., 2016; Sabuncuoğlu and Gülay, 2016; Saxena and Khanna, 2013; Shareef et al., 2019; Ünal et al., 2011; Zafar and Khan, 2011). Furthermore, studies (Tsang et al., 2004; Wang et al., 2009; Ling et al., 2010; Ünal et al., 2011 and Zafar and Khan, 2011) found that consumers had a positive attitude towards informative advertisements. However, Aksoy and Gür (2018) argued that informative advertisements published on social media had no significant impact on purchasing intent.

*Deceptiveness*

With the increase in online shopping, the frequency of seeing deceptive advertisements has increased (Toros, 2021). Deceptive advertisements can mislead consumers at a cognitive level (Olson and Dover, 1978). Deceptive advertisements make customers doubt the quality of a product. This reduces the credibility of the advertisement and diminishes its power of persuasion. Consumers who find themselves in such situation raise their defensive barriers when they come across advertisements of the product, which is continued in similar advertisements which is later seen. Companies that mostly rely on the power of advertisements for the sale of their products and services should therefore just not act to save the moment. Businesses should refrain from making deceptive claims about their products and services to maintain customer trust. Customers simply need to know the benefits of the product or service (Çakır and Çakır, 2007). According to Ducoffe (1995), advertisements perceived as deceptive have a negative effect on advertising value, and Çakır and Çakır (2007) argue that deceptive advertisements have no influence on advertising value. On the other hand, Elden et al., (2014) have shown that the perception of an advertisement as deceptive has a positive effect on advertising value.

*Entertainment*

It matters that consumers are amused by an advertisement to keep watching it. Entertaining advertisements, therefore, have a positive effect on brand attitudes (Ducoffe, 1996). This is due to the entertainment component being a factor that increases the impact of advertising, by establishing a psychological bond between the consumer and the brand (Amjad et al., 2015). Occasionally, users mention that online advertisements are more entertaining and motivating than websites themselves (Wolin et al., 2002). Studies on the consumer perception of advertisements have shown that entertaining advertisements also have a positive influence on advertising value (Bayrak Meydanoğlu, 2016; Brackett and Carr, 2001; Çakır and Çakır, 2007; Dehghani et al., 2016; Ducoffe, 1995; Elden et al., 2014; Logan et al., 2012; Murillo et al., 2016; Sabuncuoğlu and Gülay, 2016; Martins et al., 2018; Erkan et al., 2019; Saxena and Khanna, 2013; Shareef et al., 2019). At the same time, some studies (Ling et al., 2010; Tsang et al., 2004; Ünal et al., 2011; Zafar and Khan, 2011) have shown that entertaining advertisements have a positive effect on the attitude of individuals towards advertising. On the other hand, Wang et al., (2009) did not find any indications that the entertaining messages in advertising on the Internet have a positive impact on attitudes towards advertising.

*Irritation*

An irritating advertisement reduces the potential impact in the eyes of the consumer (Aaker and Bruzzone, 1985). Consumers who get irritated with an advertisement are reportedly less inclined to be persuaded by its message (Çakır and Çakır, 2007). Advertisements designed using disturbing or manipulative advertising techniques, therefore, cause negative perceptions.

Especially in recent years, there has been a proliferation of advertisements that just distract people and confuse them in mobile environments. Most of the time, this causes someone to develop a negative attitude towards the advertisement (Bayrak Meydanoğlu, 2016). In that respect, there are studies that show the negative perception of irritating advertisements (Bayrak Meydanoğlu, 2016; Murillo et al., 2016).

*Brand Awareness and Brand Association*

Brand awareness is the force of the brand that makes the consumer retain it in their memory (Pappu and Quester, 2006). The more consumers a company reaches in a short period of time, the faster that brand awareness spreads (Gwinner, 1997). In addition, brand awareness is an important element of brand equity. This has a direct impact on consumers' attitudes and perceptions of brands. This sometimes can make consumers put their brand loyalty on freeze and go for different brands (Aaker, 1992).

Brand associations refer to highlighting the important characteristics of a product or service, bringing into the picture a famous person or a symbol. The way to build a strong brand is therefore by creating an identity for the brand (Gümüş et al., 2013).

Haida and Rahim (2015) argued that informative and entertaining content in social media advertisements had a positive impact on product awareness, while Kayapınar et al., (2017) found that informative and entertaining advertising had a positive impact on brand attitudes. Apart from these, Dehghani et al., (2016) studied the influence of perceptions of YouTube advertising on advertising value, as well as the influence of advertising value on brand awareness. It was found that advertising value had a positive and significant impact on brand awareness. Martins et al., (2018) studied the impact of well-designed quality advertisements for smartphones on brand awareness. According to the results of this study, well-designed smartphone advertisements and brand awareness have a positive effect on purchasing intent. Erkan et al., (2019) examined the influence of perceptions about informative and entertaining Facebook advertisements on brand image, brand awareness and brand equity. They found that the attitude towards Facebook advertisements perceived as entertaining and informative had a positive and significant effect on brand image, brand awareness and brand equity.

*Generation Y*

Known by many different names such as Digital Generation, Millennium Generation, Eco-Explosion, Future Generation and Next Generation, this generation grew up at a time when economic conditions were slightly better and there were fewer cases of conflicts and less rampant poverty. Generation Y lives in a world where communication technologies like the Internet, mobile phones and social media that enable worldwide communication are rapidly evolving (Benckendorff et al., 2010). Generation Y likes shopping and is not shy about over-consumption. They influence their friends through consumption and they are influenced by their friends through consumption. This generation is inquisitive and highly knowledgeable about things, questions everything, is aware of their social responsibilities and is difficult to satisfy. Therefore, this generation is often the subject of research studies on marketing (Aydın and Tufan, 2018). Generation Y, which is much more active on social media, stands out from other generations with their skills in generating an intense amount of content and combining the content they receive from many different sources. Their favorite activities are status updates, checking photos shared by friends and sharing their own photos. In addition, they do not hesitate to make contact if they experience any problems with the service received from brands or companies.

*Advertising on Social Media*

Today, the average time which people spend on mobile phone internet a day is 3 hours 39 minutes, while it is 3 hours 24 minutes for TV (We Are Social, 2022). Social media advertising, another sort of Internet advertising, also makes it easier for businesses to reach consumers (Aslam and Karjaluoto, 2017). These advertisements, which classify consumers based on important personal information such as age, gender, occupation, personal interests and geographic location, quickly reach the target audience (Dehghani and Tumer, 2015).

Many studies with different perspectives on social media advertising have been conducted. Grimes (2008) has shown that the advertising effect increases with more time spent on Facebook by users. Chang (2014) noted that companies and consumers create brand value on social media by sharing photos on Instagram. Celebi (2015) found in his study on young Facebook users that the popularity of Facebook had a positive impact on behaviors and attitudes towards website advertisements. Keskin and Baş (2015) argued that consumer behavior is influenced by social media tools. According to the results of this study, consumers attach particular importance to user comments. Lee and Hong (2016) emphasized the significance of the prediction of positive user reactions to social media advertisements on Facebook. Aslan and Ünlü (2016) argued that the majority of Instagram celebrities posting social media advertisements are young, educated and single women. Lin and Kim (2016) studied user responses to unauthorized Facebook advertisements on social media. Thornhill et al., (2017) analyzed the importance of Facebook advertisements for competing brands. Zengin and Zengin (2017) studied the avoidance of Instagram advertisements by adolescents and concluded that it was more common among men. Tran (2017) found that the perception of personalized advertisements on Facebook plays an important role in increasing customer response.

## Hypotheses development

The research model developed during the literature review and tested during the study is shown in Figure 1. In this section, variables related to the described research model and hypotheses based on related literature were developed.

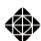

**Figure 1.** Research Model

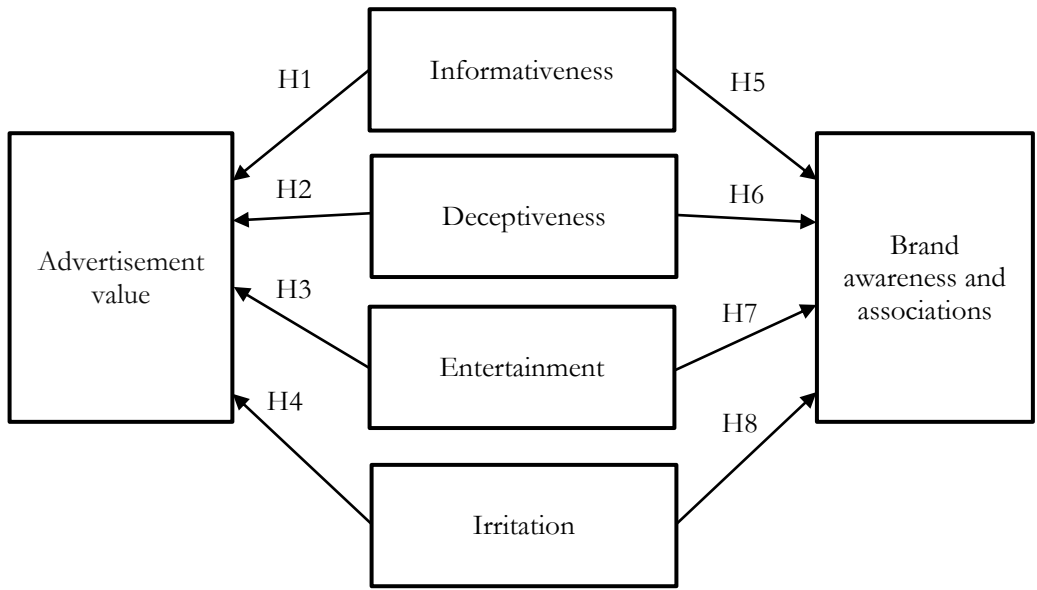

*Impact of informativeness in advertisements*

The informativeness of advertisements can influence customer satisfaction and purchasing decisions (Ducoffe, 1996). In addition, informative content of advertisements enables customers to evaluate products more rationally. Informative messages make it even more easier to access high-quality products at low prices. The informativeness of internet advertisements also helps create awareness (Saxena and Khanna, 2013). Informativeness can affect customers' buying decisions and satisfaction levels (Ünal et al., 2011). Advertisements that are made in the early days of a new product to show that it differs from its counterparts fulfill the informational function. In some areas where costs associated with reaching consumers are high, informative advertisements reduce costs and increase demand. Therefore, providing the information is an important function for adverts (Akkaya et al., 2017). Informative advertising provides the consumer with up-to-date and useful information about the product or service. In addition, the information provided in advertisements help make them more popular with consumers (Bayrak Meydanoğlu, 2016). Therefore, informative advertisements are more convincing and interesting. For this reason, consumers watch advertisements carefully to get any information that might interest them (Aaker and Norris, 1982). Consumers view advertisements more carefully to get the information they need, and give more positive reactions. This makes it easier for them to remember advertisements that they think are informative (Mehta, 2000). So the informativeness of advertisements gives brands a stronger position on the market and promotes competition (Çakır and Çakır, 2007). Advertising can also create brand awareness while being informative. It gives consumers the impression that it is the most quality product available (Çekiç Akyol, 2011). On that basis, the first research hypothesis for testing the impact of perception of informative Instagram advertisements on advertising value is as follows:

H₁. The perception of an advertisement on Instagram as informative has a positive effect on advertising value.

*Impact of deceptiveness in advertisements*

In a globalizing economy, companies compete with one another to sell their products to consumers and advertising is characterized by the competitive platforms which are popular among people. It is inevitable that advertisements are not impartial as it is the advertisers' job to convince people to buy a product or service rather than to only provide information. This in itself makes deception a common phenomenon in advertising (Toros, 2018). Consequently, deceptive advertisements designed to attract consumers' attention in the challenging competition environment may cause confusion in consumers' minds. So, a company may choose to create awareness about a product or service rather than going down the road of deception. Mass communication tools must be used correctly when doing so. This will uncover the superior aspects of the product and make the business feel more confident about advertising ethics (Karabaş, 2013). On that basis, the second research hypothesis for testing the impact of perception of deceptive Instagram advertisements on the advertising value is as follows, as determined in the literature review:

H₂. The perception of an advertisement on Instagram as deceptive has a negative effect on advertising value.

*Impact of entertaining components in advertisements*

Entertaining advertisements lead to an escape from reality, bringing aesthetic feelings to the foreground (Aydın, 2017). Therefore, it plays an important role in buying decisions when consumers enjoy online shopping (Kim et al., 2010). The entertainment aspect of advertisements refers in one respect to classic advertising content, as entertaining adverts positively impact customer loyalty. Therefore, advertisements with entertainment features are more easily accepted by customers making them more effective (Sinkovics et al., 2012). When the advertiser finally attracts the attention of the target group, it becomes even more difficult to maintain this attention. A humorous approach is one of the most effective methods to attract the attention of the audience. Humor is effective at attracting and maintaining attention. In general, consumers enjoy advertisements that entertain them and make them laugh (Hancı, 2016). The concept of 'advertainment', which is a combination of the words advertising and entertainment, is often used by advertising agencies, originating in the digital media and manifesting itself also in physical environments (Yıldırım, 2017). Entertainment also creates a feeling of satisfaction about the message of an advertisement. It plays an important role in the feelings and attitudes of consumers towards advertisements. In particular, advertisement content that is fun for the consumer has a positive effect on customer loyalty. An entertaining advertisement is perceived more positively and this has a positive effect on advertising value (Bayrak Meydanoğlu, 2016). Being entertaining allows advertisements to stand out from others. In addition, consumers send advertisements that they find entertaining to their friends and share them on the Internet (Çakır and Çakır, 2007). Making advertisements entertaining enough to arouse consumer interest makes it easier to reach the target audience. Accordingly, printed adverts are more informative while TV adverts are more entertaining (Ducoffe, 1996). On that basis, the third research hypothesis for testing the impact of perception of entertaining Instagram advertisements on advertising value is as follows, as determined based on the theoretical foundations of the study:

H₃. The perception of an advertisement on Instagram as entertaining has a positive effect on advertising value.

### The impact of irritation in advertisements

Consumers should not find an advertisement irritating or dishonest as they may think the product is not as good as what the advertisement would enable them to believe (Mehta, 2000). Irritating adverts reportedly have a negative impact on consumers (Tsang et al., 2004). If users find an advertisement confusing and distracting, the advertisement is irritating for them (Ducoffe, 1996). There is also a negative correlation between informative and irritating adverts (Aaker and Norris, 1982). On that basis, the fourth research hypothesis for testing the impact of perception of irritating Instagram advertisements on advertising value is as follows:

H₄. The perception of an advertisement on Instagram as irritating has a negative effect on advertising value.

### Impact on brand awareness and brand association

Brand awareness reflects different consumer perceptions of the brand. This awareness has its own levels. These levels are: recognition, remembrance, retention in memory, brand dominance, brand knowledge and brand concept. Brand awareness is a dimension of brand equity like perceived quality, brand associations and brand loyalty (Aaker, 1992). It also refers to consumers retaining the brand in their memories (Cop and Baş, 2010). This awareness is the entirety of impressions a consumer has in their mind about a brand. In terms of content, it includes brand recognition and brand memorability (Uztuğ, 2003). Recognition refers to whether a brand has been heard before or not, whereas memorability refers to remembering a previously encountered brand (Yılmaz, 2007).  Thus, it is being aware of a brand (Kurtbaş and Barut, 2010). Recognition is based upon information from previous encounters with the brand. This information is an acquaintance factor and is often an obstacle for better-known competitor (Gümüş et al., 2013).

From another perspective, it represents the first encounter between a brand and the consumer, i.e. the first moment of recognition. In this sense, it is very important that the brand leaves a good impression on the consumer after the first encounter. For this reason, brands must always be ready to be recognized. On the other hand, this concept, which is referred to as the sum of brand-related associations in the consumer's memory, is intended to remind the consumer of a specific product offered by the brand. This is linked to the brand awareness level of consumers. It starts with, at the lowest level, a brand going unnoticed, moving all the way up to the highest level where it is the first brand that comes to mind amongst top competitors. In such a case, consumers are more likely to choose a familiar brand than an unfamiliar one. It is therefore much more unlikely for consumers to choose a brand with a lower degree of recognition. Consumers tend to choose a brand they remember more than a brand they do not remember at all (Avcılar, 2008). In addition, consumers tend to choose the first brand that comes to mind when making a purchasing decision. Therefore, in their choices, the consumer attaches importance to the web of associations that a brand evokes in their mind (Franzen, 2002). From another point of view, it is not necessary for the consumer to remember the brand's name immediately in order to create brand awareness. There may be other things that remind the consumer of the brand. For example, a consumer may remember the shape of the packaging of a brand, the company sponsored by the brand or the advertisement slogan. An example of this is a conical perfume or a supermarket with a kangaroo logo (Lembet, 2006).

For marketing experts, brand associations facilitate the development of positive attitudes towards the brand. This is achieved through brand positioning and differentiation (Yener, 2013). Brand associations, sometimes used as brand connections, details what the brand means to consumers. It is a combination of brand identity and company connections. Brand identity refers to certain human characteristics associated with the brand. When customers choose a brand, the brand becomes meaningful with the positive emotions that arise in them (Alemdar and Dirik, 2016). Brand associations are therefore a combination of things that appear in a consumer's mind concerning the brand. It also helps the brand stand out from competitors. For example, if a person who is very thirsty first recalls to mind the cola brand, the brand association can be considered to be high. From this point of view, brand associations help consumers to process information about the brand and remember it. This has a positive effect on the consumer who finally decides to purchase the product (Koçoğlu and Aksoy, 2016).

Other research hypotheses to test whether the perception of Instagram advertisements influences brand awareness and brand association are as follows on the basis of literature review:

$H_5$. The perception of an advertisement on Instagram as informative has a positive effect on brand association and brand awareness.

$H_6$. The perception of an advertisement on Instagram as deceptive has a negative effect on brand association and brand awareness.

$H_7$. The perception of an advertisement on Instagram as entertaining has a positive effect on brand association and brand awareness.

$H_8$. The perception of an advertisement on Instagram as irritating has a negative effect on brand association and brand awareness.

## Methodology

To measure the advertising value, a study developed by Ducoffe (1995) and adapted to Turkish by Çakır and Çakır (2007) was used. Furthermore, a study developed by Yoo and Donthu (2001) and adapted to Turkish by Alemdar and Dirik (2016) was used to measure brand awareness and brand associations. Accordingly, in the first part of the survey there are 9 questions regarding demographics and the use of social media. In the second part, 19 questions were asked to measure advertising value, brand awareness and brand associations.

All statements in part two have a 5-point Likert type design. The study population is made up of Instagram users from members of Generation Y who live in the province of Gaziantep in Turkey. A convenience sampling method was used in the study, a technique not based on probability. With this technique, among those identified as a sample, only accessible persons are included in the sample (Gürbüz and Şahin, 2016). In the study, analyses were performed with a total of 665 questionnaires. The surveys were evaluated on the basis of the data obtained from the research and the statistical tests were analyzed using the SPSS and AMOS package programs.

## Results

The sample comprised 665 individuals, in which 43% (288) are male and 57% (377) are female. The average age is 32, the youngest respondent being 18 and the oldest 40 as shown in Table 1.

**Table 1.** Demographic characteristics of the participants.

|  | Frequency | Percentage |
|---|---|---|
| **Gender** | | |
| Male | 288 | 43.3 |
| Female | 377 | 56.7 |
| **Age** | | |
| 18-22 | 326 | 49 |
| 23-27 | 121 | 18.2 |
| 28-32 | 142 | 21.4 |
| 33-37 | 43 | 6.5 |
| 38-40 | 33 | 5 |
| **Marital Status** | | |
| Single | 480 | 72.2 |
| Married | 185 | 27.8 |
| **Education** | | |
| Primary | 23 | 3.5 |
| High school | 82 | 12.3 |
| Associate | 199 | 29.9 |
| Graduate | 309 | 46.5 |
| Post graduate | 28 | 4.2 |
| Doctorate | 24 | 3.6 |

The majority of youths log into Instagram more than 13 times in one day. On the other hand there are too many members of Generation Y who do not pay attention to how much time they spend on Instagram. In addition, most Generation Y users of Instagram follow personal hobby and entertainment pages. Finally the second largest platform to find members of Generation Y is YouTube as shown in Table 2.

**Table 2.** Instagram and social media habits of participants.

|  | Frequency of Females | Percentage of Females | Frequency of Males | Percentage of Males |
|---|---|---|---|---|
| **Number of Instagram Log-Ins** | | | | |
| 1-3 | 43 | 51,19 | 41 | 48,81 |
| 4-6 | 77 | 62,60 | 46 | 37,40 |
| 7-9 | 51 | 60,71 | 33 | 39,29 |
| 10-12 | 47 | 51,09 | 45 | 48,91 |
| 13 and over | 159 | 56,38 | 123 | 43,62 |
| **Length of Instagram Visit** | | | | |
| 1-3 min. | 59 | 62,11 | 36 | 37,89 |
| 4-6 min. | 79 | 59,85 | 53 | 40,15 |
| 7-9 min. | 55 | 68,75 | 25 | 31,25 |
| 10-12 min. | 62 | 55,86 | 49 | 44,14 |
| over 13 minutes | 52 | 52,53 | 47 | 47,47 |
| Do not count how much time I spend | 70 | 47,30 | 78 | 52,70 |

| Reasons for Using Instagram | | | | |
|---|---|---|---|---|
| Sharing photos or videos | 64 | 58,72 | 45 | 41,28 |
| Following brands | 28 | 50,00 | 28 | 50,00 |
| Following friends | 78 | 41,71 | 109 | 58,29 |
| Following celebrities | 16 | 48,48 | 17 | 51,52 |
| Following personal hobby and entertainment pages | 160 | 57,14 | 120 | 42,86 |
| **The Most Frequently Used Platform besides Instagram** | | | | |
| Facebook | 78 | 70,27 | 33 | 29,73 |
| Twitter | 64 | 47,41 | 71 | 52,59 |
| Google | 25 | 48,08 | 27 | 51,92 |
| YouTube | 183 | 61,41 | 115 | 38,59 |
| Pinterest | 4 | 33,33 | 8 | 66,67 |
| Snapchat | 18 | 38,30 | 29 | 61,70 |
| LinkedIn | 5 | 50,00 | 5 | 50,00 |
| **How Often Instagram Advertisements are Seen?** | | | | |
| Never | 18 | 64,29 | 10 | 35,71 |
| Occasionally | 43 | 57,33 | 32 | 42,67 |
| Sometimes | 95 | 52,20 | 87 | 47,80 |
| Often | 114 | 58,46 | 81 | 41,54 |
| All the time | 107 | 57,84 | 78 | 42,16 |

## Measurement model

To evaluate measurement model, reliability and validity of the scale were assessed. The researchers followed the two step approach suggested by Gerbing and Anderson (1988) for the measurement model construction and testing. First, the measurement model to test the reliability was examined. AVE (Average Variance Extracted), CR (Composite Reliability) and Cronbach's Alpha reliability analyses were applied to the scale to determine whether the statements on the scale were consistent.

The results are illustrated in Table 3. Cronbach's alpha of all constructs were found greater than the threshold of 0.7 (Kline, 2005) for basic research (Nunnally and Bernstein, 1994). All factors loadings were significant and above 0.6 (Bagozzi et al., 1991) and all Average Variance Extracted (AVE) values were above 0.5 (Ruvio et al., 2008; Fornell and Larcker, 1981), and composite reliabilities were above 0.7 (Hair et al., 2006).

**Table 3.** Standardized item loadings, AVE, CR and Alpha values.

| Variable | Item | Factor loadings CFA | AVE | CR | Cronbach's alpha |
|---|---|---|---|---|---|
| Advertisement value | Adv1 | 0.847 | 0.713 | 0.881 | 0.822 |
| | Adv2 | 0.864 | | | |
| | Adv3 | 0.821 | | | |
| Informativeness | Inf1 | 0.755 | 0.577 | 0.803 | 0.703 |
| | Inf2 | 0.763 | | | |
| | Inf3 | 0.761 | | | |
| Deceptiveness | Dec1 | 0.805 | 0.660 | 0.860 | 0.759 |
| | Dec2 | 0.856 | | | |
| | Dec3 | 0.783 | | | |
| Entertainment | Ent1 | 0.865 | 0.734 | 0.892 | 0.873 |
| | Ent2 | 0.868 | | | |
| | Ent3 | 0.837 | | | |
| Irritation | Irr1 | 0.871 | 0.736 | 0.848 | 0.746 |

| | Irr2 | 0.845 | | | |
|---|---|---|---|---|---|
| | Baa1 | 0.798 | 0.636 | 0.896 | 0.854 |
| | Baa2 | 0.861 | | | |
| Brand awareness-associations | Baa3 | 0.866 | | | |
| | Baa4 | 0.793 | | | |
| | Baa5 | 0.650 | | | |

Finally, we examined discriminant validity by comparing the correlations among constructs and the AVE values. As shown in Table 4, for each factor, the square root of AVE is significantly greater than its correlation coefficients with other factors, showing good discriminant validity (Gefen et al., 2000). In addition, it was examined whether the data collected from the sample had a normal distribution and the skewness and kurtosis values of the data were analyzed. If the values for skewness and kurtosis range between -2 to +2, it is assumed that the associated variable has a normal distribution (George and Mallery, 2010). The analyses showed that the data has normal distribution.

**Table 4.** The square root of AVE (shown as bold at diagonal) factor correlation coefficients.

| | Adv | Inf | Dec | Ent | Irr | Baa |
|---|---|---|---|---|---|---|
| Advertisement value | **0.844** | | | | | |
| Informativeness | 0.056 | **0.760** | | | | |
| Deceptiveness | -0.134 | 0.186 | **0.812** | | | |
| Entertainment | 0.088 | 0.497 | 0.181 | **0.857** | | |
| Irritation | 0.347 | -0.206 | -0.017 | -0.163 | **0.858** | |
| Brand awareness-associations | -0.093 | 0.242 | 0.023 | 0.122 | -0.233 | **0.797** |

**Structural model**

Structural equation analysis was conducted after satisfying the requirements of the measurement model. Table 5 list the recommended and actual values of fit indices of the normalized fit index (NFI), the comparative fit index (CFI), and the root mean square error of approximation (RMSEA). For all indices, the actual values are better than the recommended values.

**Table 5.** The recommended and actual values of fit indices.

| Fit index | chi2/df | CFI | NFI | RMSEA |
|---|---|---|---|---|
| Recommended values | <5.0* | ≥0.9** | ≥0.9*** | <0.08**** |
| Actual values | 3.889 | 0.918 | 0.902 | 0.066 |

*Bentler and Bonett (1980) **Hu and Bentler (1999)
***Fornell and Larcker (1981) ****Browne and Cudeck (1993)

In addition, Table 6 lists the path coefficients and their significance. The results in Table 6 show that informativeness ($\beta=0.178$, $p\leq0.003$) and entertainment ($\beta=0.146$, $p\leq001$) were having a positive and significant impact on advertisement value. And also informativeness ($\beta=0.254$, $p\leq001$) and entertainment ($\beta=0.144$, $p\leq001$) were having a positive and significant impact on brand awareness-associations. Therefore, H1, H3, H5 and H7 were supported. In addition deceptiveness ($\beta=(-0.263)$, $p\leq001$) and irritation ($\beta=(-0.174)$, $p\leq001$) were having a negative and significant impact on brand awareness-associations. Hence, H2 and H8 were supported. However, impact of deceptiveness on brand awareness-associations was insignificant and H6 was not supported. On the other hand impact of irritation on advertisement value was significant but it was positive and H4 was not supported as shown in Table 6.

**Table 6.** Hypothesis results.

| Structural paths | Hypothesis | Estimates | Supported or not |
|---|---|---|---|
| Informativeness → Advertisement value | H1 | 0.178* | Yes |
| Deceptiveness → Advertisement value | H2 | -0.263* | Yes |
| Entertainment → Advertisement value | H3 | 0.146* | Yes |
| Irritation → Advertisement value | H4 | 0.434 | No |
| Informativeness → Brand awareness-associations | H5 | 0.254* | Yes |
| Deceptiveness →Brand awareness-associations | H6 | -0.001 | No |
| Entertainment →Brand awareness-associations | H7 | 0.144* | Yes |
| Irritation →Brand awareness-associations | H8 | -0.174* | Yes |

**Notes:** *p < 0.05

## Discussion

Social media, a network where people share with one another their knowledge and thoughts, and communicate via the Internet, has been an important medium for the marketing world in recent years. Instagram has become the fastest growing social media environment. Publishing advertisements in these environments facilitates communication between businesses and consumers, and offers significant benefits for both businesses and consumers. However, the success of these advertisements and the degree to which they create awareness depend on the viewer's perceptions. In this context, the objective of this study is to examine whether the perception of Instagram advertisements influences advertising value, brand awareness and brand associations among Generation Y Instagram users. In this study, a research model was developed to examine the factors influencing advertising value and brand awareness. This proposed model was tested with a structural equality model. The data received from the participants was analyzed and the results interpreted.

*Theoretical implications*

Today, many companies carrying out marketing activities consider social media an important platform. Companies that maintain an effective social media management can reach their target market faster and influence their customers more easily. One of the most effective ways that companies use social media is for their social media advertising. Successful adverts increase company value in the eyes of the customers and give the brand a positive perspective. In this study conducted on that basis, it was determined whether the perceptions about Instagram advertisements, which are increasingly used in social media, influence advertising value, brand awareness and brand associations.

According to the H1 hypothesis, the perception of advertisements on Instagram as informative by Generation Y has a positive and significant effect on advertising value. In this case, members of Generation Y can be considered to be interested in and care about Instagram advertisements that they find informative. This generation is known for being inquisitive and questioning things. It can therefore be argued that they approach informative Instagram advertisements with a sense of curiosity. It can also be argued that members of this generation find these advertisements valuable because they are willing to learn about the features of a new product advertised on Instagram. The fact that informative advertisements have a positive impact on advertising value is supported by studies conducted by Ducoffe, 1995; Brackett and Carr, 2001; Çakır and Çakır, 2007; Logan et al., 2012; Saxena and Khanna, 2013; Elden et al., 2014; Bayrak Meydanoğlu, 2016; Murillo et al., 2016; Dehghani et al., 2016;

Sabuncuoğlu and Gülay, 2016; Martins et al., 2018; Erkan et al., 2019 and Shareef et al., 2019. Moreover, some studies (Tsang et al., 2004; Wang et al., 2009; Ling et al., 2010; Ünal et al., 2011 and Zafar and Khan, 2011) found that consumers had a positive attitude towards informative advertisements. Since perception is linked with attitude, it can be said that the current research has drawn similar conclusions to these studies. On the other hand, Aksoy and Gür (2018) argued that informative advertisements published on social media had no significant impact on purchasing intent. The fact that the opposite was found in this study might have to do with various factors impacting purchasing decisions.

According to the H2 hypothesis, the perception of advertisements on Instagram as deceptive by Generation Y has a negative and significant effect on advertising value. In this respect, it can be concluded that the deceptive content in Instagram advertisements detracts from the credibility of adverts in the eyes of members of Generation Y. Moreover, since this generation is made up of self-aware and knowledgeable individuals, they can be more selective and cautious when it comes to adverts which are considered to be deceptive. From another perspective, these individuals may be skeptical about the advertisements they see on Instagram and find them worthless. The fact that advertisements perceived as deceptive have a negative influence on advertising value has also been emphasized by Ducoffe (1995). On the other hand, Elden et al., (2014) have shown that the perception of an advertisement as deceptive has a positive effect on advertising value. This may be due to the age difference of participants in both studies. Because Elden et al., (2014) conducted their studies with the participation of 7-12 year old children and 19-24 year old adolescents. Çakır and Çakır (2007) argued that deceptive advertisements have no influence on advertising value. Çakır and Çakır (2007) focused their attention on TV adverts. The difference between the results of the studies may be due to the media in which the advertisements are displayed. This is due to an individual's ability to examine social media advertisements in more detail than television advertisements.

According to the H3 hypothesis, the perception of advertisements on Instagram as entertaining by Generation Y has a positive and significant effect on advertising value. As the inclusion of humorous elements in advertising is considered positive by these individuals, they tend to find such advertisements more valuable. People also want to share entertaining advertisements with other people around them. This may make entertaining Instagram advertisements more valuable to them. Thus, entertaining advertisements can create a positive perception of the product or service. The fact that entertaining advertisements have a positive impact on advertising value is supported by studies conducted by Ducoffe, 1995; Brackett and Carr, 2001; Çakır and Çakır, 2007; Logan et al., 2012; Saxena and Khanna, 2013; Elden et al., 2014; Bayrak Meydanoğlu, 2016; Murillo et al., 2016; Dehghani et al., 2016; Sabuncuoğlu and Gülay, 2016; Martins et al., 2018; Erkan et al., 2019 and Shareef et al., 2019. At the same time, some studies (Tsang et al., 2004; Ling et al., 2010; Ünal et al., 2011; Zafar and Khan, 2011) have shown that entertaining advertisements have a positive effect on the attitude of individuals towards advertising. In addition, Kim and Ko (2012) argued in their study of luxury brands on social media that entertainment factors in social media marketing have a positive effect on consumers' tendency to buy a product. On the other hand, Kayapınar et al., (2017) found that entertaining advertising had a positive and significant impact on brand attitudes. In addition, Aksoy and Gür (2018) found a positive correlation between fun social media advertisements and purchasing intent. On the other hand, Wang et al., (2009) did not find any indications that entertaining messages in advertising on the Internet have a positive impact on attitudes towards advertising. It is assumed that the reason for this is due to the difference between social media advertising and advertising on the Internet. This is because people who browse a website or are busy doing a job come across Internet advertisements.

According to the H4 hypothesis, the perception of advertisements on Instagram as irritating by Generation Y have a negative and significant effect on advertising value. However, the study has found that irritating advertisements on Instagram has a positive and significant effect on advertising value. It is assumed that people find these advertisements valuable because they get curious and want to watch it to the end, even if they find it disturbing. This is supported by studies conducted by Bayrak Meydanoğlu (2016) and Murillo et al., (2016). Bayrak Meydanoğlu (2016) studied perceptions of QR-coded advertisements in his research. They found that the perception of QR-coded mobile advertisements as irritating had a positive and significant effect on advertising value. In addition, Murillo et al., (2016) studied the effect of perceptions about Twitter advertisements on advertising value. According to study results, on the basis of the most-clicked advertisements on Twitter, being irritating has a positive impact on advertising value. Moreover, Çakır and Çakır (2007), Logan et al., (2012) and Dehghani et al., (2016) found no significant impact of irritating perception on advertising value. On the other hand, Ducoffe, 1995; Saxena and Khanna, 2013; Elden et al., 2014; Dehghani et al., 2016; Martins et al., 2018; Erkan et al., 2019; Shareef et al., 2019 found that perceptions of an advertisement as irritating will have a negative and significant impact on advertising value.

According to the hypotheses H5 and H7, the perception of advertisements on Instagram as informative and entertaining has a positive and significant effect on brand awareness and brand associations. When members of Generation Y perceive Instagram advertisements as informative and entertaining, these advertisements raise awareness of the related product or service. It can be argued that such people find advertisements that they consider entertaining and informative useful and carefully examine the brand of the advertised product or service. These findings were compared with previous studies on similar topics. Haida and Rahim (2015) and Kayapınar et al., (2017) discovered similar findings. Haida and Rahim (2015) argued that informative and entertaining content in social media advertisements had a positive impact on product awareness. Kayapınar et al., (2017) found that informative and entertaining advertising had a positive impact on brand attitudes. Apart from these, Dehghani et al., (2016) studied the influence of perceptions regarding YouTube advertising on advertising value, as well as the influence of advertising value on brand awareness. It was found that advertising value had a positive and significant impact on brand awareness. Martins et al., (2018) studied the impact of well-designed quality advertisements for smartphones on brand awareness. According to the results of this study, well-designed smartphone advertisements and brand awareness have a positive effect on purchasing intent. Erkan et al., (2019) examined the influence of perceptions about informative and entertaining Facebook advertisements on brand image, brand awareness and brand equity. They found that the attitude towards Facebook advertisements perceived as entertaining and informative had a positive and significant effect on brand image, brand awareness and brand equity.

According to the H8 hypothesis, the perception of advertisements on Instagram as irritating by Generation Y has a negative and significant effect on brand awareness and brand associations. This is a finding that is also supported by studies conducted by Haida and Rahim (2015) and Dehghani et al., (2016). Haida and Rahim (2015) stated that the perception of social media advertisements as irritating affects product awareness negatively. In a study on YouTube advertisements, Dehghani et al., (2016) found that using irritating elements in an advertisement has a negative effect on advertising value, which in turn has a negative impact on brand awareness. In addition, Tsang et al., (2004) argued that advertisements shown without permission have a negative impact on the brand. Hutter et al., (2013) stated that the discomfort felt about the Facebook fan page has a negative impact on WOM.

*Practical implications*

According to the findings of this study, the perception of advertisements on Instagram as informative by Generation Y has a positive effect on advertising value. Advertising agencies, business owners and executives who want to appeal to the younger generation through Instagram advertisements should focus on informative content to improve the perception of their advertisements. Advertising value is a measure of the usefulness and success of advertising. Interestingly designed advertisements with compelling content and accurate information will increase the value of Instagram advertisements. Advertising agencies, business owners and executives should therefore take this into account and prepare content accordingly. In the end, one of the most important functions of advertising is to inform the consumer about products or services. Therefore it has a positive effect on the advertising value if new and unknown details about products are published on social media. Creating informative content on Instagram can be done faster through story advertisements or collection advertisements. Indeed, the first image that is intended to convey information should be clearly obvious and direct. Furthermore, companies that wish to appeal to Generation Y should decorate messages with an interesting title without making it too obvious that it is an advertisement. That way, while the advertisement conveys a useful message, the company or brand can be advertised in the background.

Also, the study revealed that deceptive advertisements on Instagram have a negative impact on advertising value from the point of view of Generation Y. This is an indication that some advertisements on Instagram are only sales-oriented. Such advertisements can even cause the product to be perceived as if they are guaranteed by Instagram. For example, the phrase "Revitalizes Body and Brain" used to promote an energy drink containing high amounts of caffeine and sugar has long been the subject of complaints on social media. In that respect, advertising agencies, business owners and executives are advised to provide accurate information on Instagram advertisements and to avoid misleading content.

Moreover, according to the findings of this study, the perception of advertisements on Instagram as entertaining by Generation Y has a positive effect on the perceived advertising value. The purpose of using entertaining content in advertising is to promote products and services by entertaining consumers rather than conveying the intended message directly. So, advertising agencies, business owners and executives should be focusing on providing entertaining content to increase the advertising value for Instagram advertisements. For this reason, in many popular television commercials, the entertainment is delivered by popular comedians. Similarly, using famous comedic figures on social media advertisements can increase the advertising value of advertisements on Instagram. In addition, entertainment can be brought to the fore through competitions or digital games in Instagram advertisements. It is now possible to create more colorful and attractive advertisements through technologically advanced visual and acoustic elements. Instagram also comes with some handy features to make advertisements more entertaining. However, it will be more efficient to prepare it more professionally, or to have it prepared by graphic designers or designers to increase the advertising value. It will also be useful to use videos to publish fun advertisements on Instagram. But the most important thing to pay attention to is the first few seconds of the video. This is because videos that do not get the viewer's attention in the first few seconds tend to get ignored. Some viewers even watch videos in a quiet environment. Advertising agencies, business owners, and executives are encouraged to consider these points when creating entertaining content on Instagram.

According to other findings, the perception of advertisements on Instagram as informative and entertaining by Generation Y has a positive effect on brand awareness and brand associations. Brands have different purposes when they advertise on Instagram. Foremost among them are; creating brand awareness, increasing sales or winning followers. However, it is not easy to build brand awareness. According to the current research, this can be done through entertaining and informative Instagram advertisements. Accordingly, advertising agencies, business owners and executives may be advised to give greater weight to informative and entertaining advertising in order to positively influence brand perception. On Instagram, writing informative content about the brand in the description area will get users' interest and allow them to learn new things. With this content, users will be more likely to leave comments and increase their contact with the brand. This will make people who are exposed to the advertisement fully aware of the brand. Using informative hashtags to ensure brand awareness is another good step that can be taken. If hashtags are used, people can communicate more easily with the brand. On the other hand, making use of professionals and experienced social media specialists who use visual and humorous elements to create entertaining advertisements will have a positive impact on brand awareness and association.

The majority of youths from Generation Y log into Instagram more than 13 times in one day. It is therefore advisable for advertising agencies, business owners and executives to take this into account and keep their advertisements up to date at regular intervals throughout the day. This is due to users wanting to see different content every time they log into Instagram during the day. In this context, it is advisable for advertising agencies, business owners and executives to have advertisers create and present more than one content at different times of the day. On the other hand, there are too many members of Generation Y who do not pay attention to how much time they spend on Instagram. This way, users will be more impressed by Instagram advertisements with engaging content. To achieve this, it is advisable to provide several visual contents at the same time. On Instagram, this can easily be done with story advertisements, rotating advertisements, and collection advertisements. In addition, most Generation Y users of Instagram follow personal hobbies and entertainment pages. In view of this, advertisers may be advised to place advertisements through people who like hobby pages or funny pages. Besides, the second largest platform to find members of Generation Y is YouTube. Therefore, advertising agencies, business owners and executives can be recommended to prepare advertisements for YouTube.

## Conclusion

The study was conducted to demonstrate the impact on value perception and awareness towards advertisements using the theoretical framework of Ducoffe's (1995) advertising value model. This study provides insights into the importance of selecting the right content to increase advertising value and brand awareness of Instagram advertisements. The aim is to conceptualize consumers' perceptions of Instagram advertisements. This study examined the impact of consumer perceptions of Instagram advertisements on advertising value, brand awareness and brand association from the point of view of members of Generation Y. According to the results, informative and entertaining Instagram advertisements have a positive impact on advertising value, while deceptive advertisements have a negative impact on advertising value. In addition, informative and entertaining advertisements have a positive impact on brand awareness and brand associations while irritating advertisements have the opposite impact. Participants were also asked about their social media and Instagram habits. On the basis of the answers, the majority of youths from Generation Y log into Instagram more than 13 times in one day. Accordingly, it can be argued that the respondents log into Instagram frequently during the day. In addition, the majority of respondents do not pay attention to how much time they spend on Instagram. So it can be assumed that members of Generation Y spend most of their time on Instagram in this environment and log into Instagram whenever they have free time. In addition, the majority of respondents use Instagram to follow personal hobbies and entertainment pages. Thus, it can be argued that members of Generation Y consider Instagram an interesting place where they can have a pleasant time. According to the results of the study, the second platform where respondents spend most of their time is YouTube. The rapid growth of YouTube in recent years has also been documented by the study. With the emergence of a new group of social media professionals called YouTubers, many members of Generations Z and Y have begun to publish their own content on YouTube and make money out of them. Finally, the majority of the respondents reported having encountered Instagram advertisements. This is an indication that Instagram advertisements have become quite popular and that people pay attention to them.

## Limitations and future research

Individuals with an Instagram account were included in the study whereas those without an Instagram account were excluded from the study. Another limitation of the study is that people from Generation Y were included in the study, while people from other age groups were excluded. The last limitation was the inclusion of individuals from the province of Gaziantep in Turkey chosen through convenience sampling and the exclusion of others that did not fit these criteria. For this reason, caution is recommended concerning the generalizability and external validity of the results of this study.

When the appropriate conditions are met, a study population made up of more Instagram users can be selected, and research can be conducted in larger areas. The results of a study conducted with a larger study population and sample will provide researchers with deeper insights into the topic. Furthermore, demographic data in the study is limited. Further demographic data such as income status and occupation may be collected in future studies. This way, it can be examined if the findings differ when such variables are added.

The study will be beneficial for researchers working in the field of social media advertising. Future studies can compare advertisements in different social media environments using different age groups. For instance, studies on Twitter advertisements can be conducted for Generation X while the same can be done on YouTube advertisements for Generation Z.

Finally, researchers can study advertisements promoted by social media celebrities. Such a study can give researchers an insight into the usefulness of social media advertisements that employ internet celebrities. In this context, the results of such a study can guide researchers who want to work in the field of social media marketing.